\documentclass[preprint2,twoside]{aastex}

\usepackage{graphics}
\usepackage{psfig}

\def\R{{\sl ROSAT}}
\def\A{{\sl ASCA}}
\def\C{{\sl Chandra}}
\def\E{{\sl Einstein}}
\def\B{{\sl BeppoSAX}}
\newcommand\hii{H{\small II}}
\newcommand\hi{H{\small I}}
\def\cor{\widehat=}
\def\gs{\mathrel{\mathchoice {\vcenter{\offinterlineskip\halign{\hfil
$\displaystyle##$\hfil\cr>\cr\sim\cr}}}
{\vcenter{\offinterlineskip\halign{\hfil$\textstyle##$\hfil\cr
>\cr\sim\cr}}}
{\vcenter{\offinterlineskip\halign{\hfil$\scriptstyle##$\hfil\cr
>\cr\sim\cr}}}
{\vcenter{\offinterlineskip\halign{\hfil$\scriptscriptstyle##$\hfil\cr
>\cr\sim\cr}}}}}
\def\ls{\mathrel{\mathchoice {\vcenter{\offinterlineskip\halign{\hfil
$\displaystyle##$\hfil\cr<\cr\sim\cr}}}
{\vcenter{\offinterlineskip\halign{\hfil$\textstyle##$\hfil\cr
<\cr\sim\cr}}}
{\vcenter{\offinterlineskip\halign{\hfil$\scriptstyle##$\hfil\cr
<\cr\sim\cr}}}
{\vcenter{\offinterlineskip\halign{\hfil$\scriptscriptstyle##$\hfil\cr
<\cr\sim\cr}}}}}

\begin{document}

\title{\bf \R\ X-Ray Observations of the Spiral Galaxy M81}

\author{Stefan Immler \& Q. Daniel Wang}
\affil{Astronomy Department, University of Massachusetts, Amherst, MA 01003}
\affil{immler@astro.umass.edu \& wqd@astro.umass.edu}

\shorttitle{\R\ X-ray Observations of M81}
\shortauthors{Immler \& Wang}

\begin{abstract}
We present results from the analysis of deep \R\ HRI and PSPC observations of
the spiral galaxy M81. The inferred total (0.5--2~keV band) luminosity 
of M81 is $\sim3 \times 10^{40}$ ergs s$^{-1}$, excluding the contribution
from identified interlopers found within the $D_{25}$ ellipse. 
The nucleus of the galaxy alone accounts for about $65\%$ of this 
luminosity. The rest is due to 26 other X-ray sources (contributing 
$\sim10\%$) and to apparently diffuse emission, which is seen across much of the 
galactic disk and is particularly bright in the bulge region around the nucleus.
Spectral analysis further gives evidence for a soft component, which can be characterized 
by a two-temperature optically thin plasma with temperature at $\sim 0.15$~keV and 0.60~keV 
and an absorption of the Galactic foreground only. These components, accounting for $\sim13\%$ of 
the X-ray emission from the region, apparently arise in a combination of hot gas and faint 
discrete sources. We find interesting spatial coincidences of luminous 
($10^{37}$--$10^{40}~{\rm ergs~s}^{-1}$)
and variable X-ray sources with shock-heated optical nebulae. Three of them
are previously classified as supernova remnant candidates. The other one 
is far off the main body of M81, but is apparently associated with a dense \hi\ 
concentration produced most likely by the tidal interactions of the galaxy with 
its companions. These associations suggest that such optical nebulae
may be powered by outflows from luminous X-ray binaries, 
which are comparable to, or more luminous than, Galactic `micro-quasars'. 

\end{abstract}

\keywords{galaxies: general---galaxies: spiral---X-rays: general---X-rays: galaxies---X-rays: ISM}

\section{Introduction}
\label{introduction}

`Normal' galaxies of all morphological types are known to be X-ray emitters,
with luminosities in the soft ($\sim0.2$--3.5~keV) X-ray band ranging from
$10^{38}$ to some $10^{41}$ ergs s$^{-1}$ (Fabbiano 1989). Despite this band being a 
relatively small part of the electromagnetic spectrum and the X-ray output
representing only a small fraction of the bolometric luminosity of a galaxy, 
X-ray observations are uniquely suited to studying various astrophysical 
phenomena elusive in other wavelength regimes. 
Whereas the overall emission of a galaxy is dominated by stars in the optical
and reprocessed stellar emission in the infrared, X-ray emission is primarily 
related to the high-energy phenomena associated with end-products of stars:
accreting compact objects, supernova remnants, and the shock-heated hot interstellar medium. 

M81 (Table~\ref{parameters}) is an ideal test-bed for our understanding of the X-ray 
emission components in early-type spirals since it is the nearest such galaxy outside 
of the Local Group. It is located in a direction with relatively low Galactic absorption
($N_{\rm H}=4\times10^{20}~{\rm cm}^{-2}$; Dickey \& Lockman 1990), considerably lower than 
that along the line-of-sight to M31 ($7\times10^{20}~{\rm cm}^{-2}$). 
Compared to the high disk inclination of M31 ($i=77\degr$), 
the moderate inclination of M81 ($i=32\degr$) makes its emission components less
obscured internally and easier to disentangle spatially.
M81 hosts a low-luminosity Seyfert nucleus with some characteristics of a low-ionization
nuclear emission-line region (LINER). The nucleus itself has been studied intensively in 
many wavelength regimes (e.g., Ishisaki et al. 1996; Kaufman et al. 1996; Ho et al. 1999 
and references therein). More recently, the outburst of SN~1993J in M81 led to a large 
number of additional observations, especially in the X-ray band (e.g. Zimmermann et al. 1994a).

Previous \E\ observations of M81 resulted in the detection of nine individual
X-ray sources including the nucleus, with luminosities in excess of 
$2 \times 10^{38}$ ergs s$^{-1}$ in the 0.2--4.0~keV band (Fabbiano 1988). 
Only marginal evidence for variability was found for the nucleus and two other 
sources. About two-thirds of the non-nuclear emission was not resolved spatially.

Spectral and temporal properties of the M81 nucleus have been studied, based on 
both \A\ and \B\ data (Ishisaki et al. 1996; Iyomoto \& Makishima 2000;
Pellegrini et al. 2000). The spectrum of the nuclear source is well described by 
a power-law with a photon index $\Gamma = 1.85$. In addition, 
a softer component with a temperature of 0.6--0.8~keV was also detected.
Long-term X-ray variability of the M81 nucleus was found, with variations up to a 
factor of three over a period of 5.5 years, together with intra-day variability of 
30\% (Iyomoto \& Makishima 2000).
The spectral properties, together with the observed high variability, were
regarded as being typical for a low-luminosity AGN.
The nature of the softer component, however, could not be specified.

This paper reports on the results of deep X-ray observations of M81 obtained
with the Position Sensitive Proportional Counter (PSPC) and the High Resolution
Imager (HRI) onboard \R\ (Tr\"umper 1983). We focus mainly on the discrete 
X-ray source population inside the galaxy and the diffuse emission from the disk
and the extended bulge of M81. After a description of the X-ray observations and 
the data calibration (\S\ref{obs}), we explain the data analysis strategies
for the detection of discrete X-ray sources, their variability
and the diffuse emission (\S\ref{analysis}).
X-ray properties of discrete sources are examined in detail in \S\ref{point_sources}, 
followed by a discussion of the most interesting sources in a multi-wavelength context in 
\S\ref{multiwavelength}. Results on the diffuse X-ray emission are presented in 
\S\ref{results_diffuse}. In \S\ref{m101}, we compare  global X-ray properties of M81 and 
the late-type spiral M101, for which a similar study has been performed.
The results and conclusions are summarized in \S\ref{summary}.

\section{Observations and Data Calibration}
\label{obs}

\R\ X-ray observations of M81 were used to construct a single 177 ks HRI 
image from 72 observation intervals (OBIs) included in 10 individual observations,
and a single 101~ks PSPC image from 54 OBIs in 8 individual observations. 
Table~\ref{obs_tab} lists individual X-ray observations used in the present study. 

Attitude solutions of \R\ pointings derived from the Standard Analysis 
Software System (SASS, Voges et al. 1992) show residual attitude errors
of $\sim 6''$. To improve the attitude solution for the high-resolution 
HRI observations (spatial resolution $\sim5''$ FWHM on-axis), we 
used five bright point-like X-ray sources with off-axis distance $<8'$ from 
the pointing direction visible in every single OBI to align each of the
different OBIs with respect to the first. Positional offsets were all
$\le6\farcs5$. A comparison between the radial intensity profiles of
the five co-added point sources before and after the attitude correction shows
that the average point spread function (PSF) of the \R\ XRT+HRI is
improved from $11''$ to $9''$ FWHM. The observations were merged with respect 
to a nominal pointing direction of R.A., Dec. (J2000) 
$=09^{\rm h} 55^{\rm m} 33\fs6$, $+69^\circ 04^\prime 16^{\prime\prime}$.

To further correct the merged HRI image for systematic pointing offsets, we compared 
centroid positions of 8 point-like X-ray sources with optical counterparts 
listed in the APM catalog (Irwin et al. 1994) and the HST Guide Star 
Catalog (\S\ref{point_sources}). A satisfactory position alignment was found 
by shifting the X-ray images 1\farcs0 to the west and 4\farcs0 to the north.

Complementing the \R\ HRI data, PSPC observations of M81 (cf. Table~\ref{obs_tab})
were used  to study spectral and timing characteristics of X-ray sources.
To match the HRI field of view, only the inner $<17'$ region of the entire $2^\circ$ 
diameter PSPC field-of-view was used. The PSPC observations cover the same
(0.1--2.4~keV) band, while having an on-axis angular resolution of $\sim 25''$ and 
an energy resolution of $\delta E/E \sim 0.43 (0.93~{\rm keV}/E)^{0.5}$.

Data reduction was performed with the EXSAS software (Zimmermann et al. 1994b) and 
with our own IDL programs. Background and exposure maps were constructed with the ESAS 
software (Snowden \& Kuntz 1998). The background subtracted and exposure corrected images
of M81 are given in Figs.~\ref{f1} and \ref{f2}.

\section{Data Analysis}
\label{analysis}

\subsection{Detection of X-Ray Sources}

For the HRI source detection, we conducted local (sliding box) and map 
detections, using images of pixel size $5''$, as well as maximum likelihood analysis of 
individual sources. In order to reduce the background due to UV emission and cosmic rays, 
only HRI PI channels 2--10 were used. Sources detected  with a likelihood of 
$L\ge8$ were accepted and merged in a source list (Table~\ref{hri_sourcelist}).
Maximum likelihood values $L$ can be converted into Gaussian probabilities 
$P$ using the relation $P = 1-e^{-L}$ (cf., Cruddace et al. 1988).
Accordingly, a likelihood of 8 corresponds to $3.6\sigma$ significance.
All sources with an off-axis angle exceeding $17'$ were excluded from the source list
due to the deteriorated PSF and the count rate uncertainties at these large off-axis 
angles. Special care was taken in regions of enhanced diffuse emission regions. 
Sources only detected by the map algorithm were not accepted in the bulge region. 
Similarly, we searched for sources in the PSPC hard band (0.5--2~keV; PI channels 52--201;
Table~\ref{pspc_sourcelist}). Sources with S/N $\ge4$ were selected. The PSPC soft band is 
dominated by the local diffuse Galactic background emission, and no additional source is detected.

Count rates may be converted into fluxes by assuming an X-ray spectral model. 
The X-ray spectra from Galactic stars, SNRs and 
hot gas observed in various emission regions within galaxies (e.g. galactic halos, 
\hii\ regions or accretion of material onto a compact object) can often be 
characterized as optically-thin thermal plasma with temperatures of a few $\times 
10^6$~K. AGNs typically show non-thermal emission, its soft X-ray band spectrum 
being described by a power-law with a photon index of $\sim2$. 
A reasonably good approximation of the conversion factor for these spectra is 
$4 \times 10^{-11}~({\rm ergs~cm}^{-2}~{\rm s}^{-1})/({\rm cts~s}^{-1})$ 
(cf. Fig. 5 in Wang et al. 1999), which is adopted throughout the paper. 
The uncertainty of this conversion factor for different source spectra is 
less than a factor of $\sim 2$, except for sources located within or beyond 
regions with foreground column densities in excess of several $\times 10^{21} {\rm~cm^{-2}}$. 
A count rate of $1\times 10^{-4}$ cts s$^{-1}$ hence converts to an unabsorbed 
(0.5--2~keV band) source flux of $f_{\rm x} \sim 4 \times 10^{-15}$ ergs cm$^{-2}$ s$^{-1}$ 
and a luminosity of $L_{\rm x} \sim 6 \times 10^{36}$ ergs s$^{-1}$, assuming a distance
of 3.6~Mpc.

For the conversion between the \R\ HRI and PSPC count rates, a factor of 3
was used. For very soft sources or sources with little absorption (e.g. unabsorbed 
foreground stars), however, the conversion factor can be significantly
higher ($\sim 8$), while the conversion factor for sources with strong 
absorption could be as low as $\sim 2.5$.

\subsection{Timing Analysis of X-Ray Sources}
\label{variability}

The \R\ data allow for time variability analysis of point sources on 
various time scales. We constructed lightcurves for all HRI sources 
(Table~\ref{hri_sourcelist}) and performed statistical tests to check for 
variability. To achieve reasonable counting statistics, we binned the data into
15 observation blocks with $\sim11$~ks exposure time each. Furthermore, PSPC 
lightcurves were constructed for the five brightest X-ray sources in the M81 field 
from 13 observation intervals with $\sim8$~ks integration time each. 
The time-dependent background was determined by normalizing the total 
background map of the HRI and PSPC images according to 
the total source-removed count rate in each exposure interval.
Background subtracted source counts were extracted within the 90\% radii
around the fixed source positions found by the source detection algorithm.
We tested the variability of each source by using a $\chi^2$ test.
To verify the $\chi^2$ test, a Kolmogorov-Smirnov test was performed 
on the unbinned lightcurves of all HRI sources. Both statistical tests resulted 
in an identical list of variable sources. The combined \R\ HRI and PSPC 
lightcurves of the five brightest sources are given in Fig.~\ref{f3}. 

\subsection{Analysis of Diffuse X-ray Emission}
\label{diffuse}

A substantial fraction of the X-ray emission from M81 cannot be accounted for by the
detected X-ray sources, but is confused by bright X-ray sources, especially the nucleus.
For an effective point-like source subtraction, we compared the radial intensity profiles 
of SN~1993J in the merged HRI image and LMC X-1 in an on-axis calibration observation 
(obs. ID 150013h; 869~s exposure time); the latter source is substantially brighter. 
The two profiles agree with each other. Furthermore, the LMC X-1 and 
the nucleus of M81 have similar X-ray spectra. Thus, we used the LMC X-1 as an on-axis 
model PSF for the M81 HRI image. Fig.~\ref{f4} compares the radial surface brightness 
profile around the nucleus of M81 with the normalized LMC X-1 profile.
Detected X-ray sources were excised, except for the central source.
The LMC X-1 profile was scaled to the peak of the M81 X-ray emission.
Background was added to match the M81 background level of
$1.5 \times 10^{-2}$ cts arcsec$^{-2}$ s$^{-1}$, estimated at a radius of $\sim 4'$.
The residual was integrated to estimate the enhancement of apparently diffuse emission 
in the bulge region of M81.

\section{Discrete Sources}
\label{point_sources}

We detect 46 HRI X-ray sources in the $17\arcmin$ radius field.
Source positions, together with the source numbers, are illustrated in Fig.~\ref{f2}.
The X-ray properties of the sources are summarized in Table~\ref{hri_sourcelist}: 
source number (col. 1, `H' denotes HRI), right ascension and declination (cols. 2 and 3), 
90\% confidence error radius of the source position (col. 4, including $3\farcs5$ 
systematic error for the attitude solution), net count rate and error, corrected for 
vignetting, exposure and dead-time (col. 5), and comments regarding the variability of 
each source (col. 6). 

Table~\ref{pspc_sourcelist} lists properties of 69 PSPC sources detected 
in the the same field: source number (col. 1, `P' denotes PSPC), right ascension 
and declination (cols. 2 and 3), net count rate and error in the 0.5--2~keV
band (col. 4), and `hardness ratios' HR1 and HR2 (cols. 5 and 6). The hardness ratios are 
defined as ${\rm HR1}=({\rm hard}-{\rm soft})/({\rm hard}+{\rm soft})$ and
${\rm HR2}=({\rm hard2}-{\rm hard1})/({\rm hard2}+{\rm hard1})$, where
`soft' and `hard' denote source net count rates in the $0.11$--$0.41~{\rm keV}$
and $0.52$--$2.01~{\rm keV}$ bands, respectively. `Hard1' and `hard2' are the split of a 
hard-band rate into two: $0.52$--$0.90~{\rm keV}$ for `hard1' and 
$0.91$--$2.01~{\rm keV}$ for `hard2'. We set ${\rm HR1}=1.0$ for several sources with 
unphysical values of ${\rm HR1}>1$, which are a result of statistical uncertainties in the 
data, most seriously in the soft band. The hardness ratios, which are sensitive to both the 
line-of-sight X-ray absorption (HR1) and the intrinsic `hardness' of the spectrum (HR2), 
are helpful for discriminating between Galactic and extragalactic sources 
(Wang et al. 1999). A source with HR1 $\lesssim 0.4$ is most likely a Galactic foreground
star, whereas a source with HR2 $\gtrsim 0.4$ is likely to be extragalactic.
Also included in Table~\ref{pspc_sourcelist} are corresponding HRI source numbers based 
on positional coincidences within the detection apertures, as well as comments on the 
variability for each individual source (col. 7).

The 0.5--2~keV band flux range for all sources found in the M81 field is
$5 \times 10^{-15}$ -- $1.2 \times 10^{-11}$ ergs cm$^{-2}$ s$^{-1}$.
Within the $D_{25}$ ellipse of M81, 27 HRI and 28 PSPC sources are detected.
Six of the PSPC sources (P20, P21, P28, P30, P32, P50) have no HRI counterpart,
whereas two PSPC sources (P41, P43) are composites of multiple HRI sources.
Excluding identified interlopers and the nuclear sources, the remaining 26 X-ray sources
detected within the $D_{25}$ ellipse of M81 (22 HRI sources and 4 additional PSPC sources) 
have fluxes ranging from $0.5$ to $78 \times 10^{-14}$ ergs cm$^{-2}$ s$^{-1}$.
The corresponding luminosity range is $8\times 10^{36}$  -- 
$1.2\times 10^{39}$ ergs s$^{-1}$.

We flagged sources with a probability of variability $\gs 0.9973$ (corresponding to
3 Gaussian sigmas) as `var' in Tables~\ref{hri_sourcelist} and \ref{pspc_sourcelist} 
(sources H9/P21, H13/P29, H15, H17/P35, H18/P36, H21/P37, H22/P38, H25, H38/P47 and H44/P66).
Each of the five brightest sources in M81 is variable with a significance $>3\sigma$ 
(Fig.~\ref{f3}). The nuclear X-ray source, for example, shows strong variability with 
amplitude up to a factor of 2.5, consistent with previous \A\ observations (Ishisaki 
et al. 1996; Iyomoto \& Makishima 2000). 

\section{Multi-Wavelength Properties of X-ray Sources}
\label{multiwavelength}

Table~\ref{ident} presents our preliminary classifications of the detected X-ray 
sources, based on their optical properties. We cross-correlate the positions
of the X-ray sources with various publically available catalogs and
identify plausible optical counterparts within twice the 90\% 
confidence radius around each X-ray source. The 
APM magnitude limit is 21.5 mag for the DSS1 blue plates and
20.0 mag for the red plate (Irwin et al. 1994). The USNO A-v2.0 Catalog 
of Astrometric Standards has limiting magnitudes of ${\rm O}=21$, ${\rm E}=20$, ${\rm J}=22$, 
and ${\rm F}=21$. As results, we classify eleven HRI and nine additional PSPC sources 
as foreground or background objects (`interlopers', i.e. AGN, quasars, 
Galactic stars, etc.). Given a limiting X-ray flux of 
$4 \times 10^{-15}$ ergs cm$^{-2}$ s$^{-1}$, the number of expected interlopers in
the $17'$ radius field is $\sim13$, using results of the \R\ Medium 
Sensitivity Survey (Hasinger et al. 1991). Thus, our identification of detected X-ray 
sources with interlopers is nearly complete and the bulk of the remaining X-ray sources 
within the $D_{25}$ ellipse of M81 is associated with the galaxy. In the following, we discuss
the most interesting sources in a multi-wavelength context.

\noindent $\bullet$ {\sl Correlation of X-ray sources with supernova remnant candidates:} \\
Of the 22 HRI sources assumed to be associated with M81, we find that three sources 
spatially coincide with SNR candidates (Table~\ref{snr}). The identification of 41 SNR 
candidates in M81 was obtained by  Matonick \& Fesen (1997; hereafter MF), 
based on measured [S{\sc ii}]/H$\alpha \gs 0.45$ ratios.
The three proposed SNR candidates in M81 have luminosities in the range
$10^{37}$--$10^{39}$~ergs~s$^{-1}$. In comparison, X-ray luminosities of 
SNRs in our Galaxy typically are in the order of $10^{35}$--$10^{37}$~ergs~s$^{-1}$.
Both M83 and M101 show significant correlations between
similarly bright X-ray sources with SNR candidates (Immler et al. 1999; Wang 1999).
For M81, the random chance coincidence of any of the 41 SNR candidates being inside the 
search radius of any of the HRI sources is $0.3$. If progenitors of SNe and X-ray sources are
spatially correlated, however, the probability for the chance coincidence will be higher. 
Nevertheless, the apparent spatial coincidence of the X-ray sources with the SNR 
candidates are intriguing. 

Interestingly, our X-ray timing analysis shows that all three X-ray sources are variable 
(Table~\ref{hri_sourcelist}). Source H9 is only detected in the first of the 15 HRI 
observation blocks used for the timing analysis. Also, source H13 is only detected in 
observations no.~1--3 and 15 while being below the detection threshold for the remaining 
epochs ($\ls 10^{37}$ ergs s$^{-1}$). The X-ray lightcurve of the bright X-ray source H21/P37
shows both long-term and short-term variability, with variations in the X-ray flux of a 
factor $\sim2$ within six days (Fig.~\ref{f3} at day 950). A PSPC spectrum, extracted from 
within a circle of $25''$ radius around the source, is well fitted by a thermal bremsstrahlung 
model with a temperature of ($1.7 \pm 0.5$)~keV and an absorbing column of 
$N_{\rm H}=(2.6 \pm 0.4) \times 10^{21}$ cm$^{-2}$, or by a power-law model with 
a photon index $\Gamma=2.4\pm0.4$ and an absorbing column of
$(3.4 \pm 0.2) \times 10^{21}$ cm$^{-2}$. An optically thin thermal plasma model does not 
give an acceptable fit. The high X-ray luminosity, the strong variability, and apparently
heavily absorbed spectral properties are the characteristics of a superluminous
X-ray binary containing a black hole primary.

Could the association between X-ray binaries and the remnants be physical?
Energetically, an X-ray binary has an energy capacity of $\sim 2 \times 
10^{53} M_{\rm c}/M_\odot$ ergs (i.e., $\sim 10\%$ of the rest mass $M_{\rm c}$ 
of the companion). A large fraction of the energy can be released as outflows as observed 
in Galactic `micro-quasars' (e.g., Mirabel \& Rodr\'{\i}guez 1994). Such outflows may well 
create ISM structures such as the remnant candidates, and even some superbubbles.

\noindent $\bullet$ {\sl H44/P66 and a superbubble:} \\
The second-brightest X-ray source (H44/P66) in the field 
positionally coincides with an identified emission-line nebula (Miller 1995)
located $\sim2'$ east of the dwarf companion Ho IX. The source was 
first detected by an \E\ observation (source X9; Fabbiano 1988), with a
flux consistent with the mean \R\ flux. A shell-like optical nebula of size 
$\sim 14''\times 27''$, corresponding to $\sim 250~{\rm pc} \times 475~{\rm pc}$ 
at the distance of M81, was later identified in the region (Miller 1995). 
The presence of strong [S{\sc ii}] and [O{\sc i}] emission lines in optical spectra 
indicate that the nebula is shock-heated. The nebula was thus classified as a superbubble. 
Our timing analysis, however, reveals strong variability with an amplitude exceeding a factor 
of 2.5 (cf. Fig.~\ref{f3}). Therefore, the X-ray emission cannot originate in diffuse hot gas.
We also find that the PSPC spectrum is reasonably well characterized  by a thermal 
bremsstrahlung spectrum with a temperature of ($1.0\pm0.1$)~keV or a power-law with index 
$\Gamma=2.1\pm0.2$. In both cases the absorbing column density
($\sim 2$ and $\sim3 \times 10^{21} {\rm~cm^{-2}}$, respectively)
is significantly higher than the expected Galactic absorption, but is consistent 
with the inclusion of the column density seen in the \hi\ map of M81 (Fig. 7; Yun et
al. 1994). This indicates that the source is within or beyond the \hi\ concentration in the 
same region. An \A\ observation of the source in the 0.5--10~keV range further shows a flat
and featureless spectrum, which will be presented in a separate publication. 
Therefore, the source is most likely an X-ray binary that contains an accreting black hole 
primary. The apparent source/nebula association then again indicates a physical connection
between a superluminous X-ray binary and a shell-like ISM structure. 

\noindent $\bullet$ {\sl Supernova 1993J:} \\
The type IIb SN~1993J is the main motive
for the wealth of the \R\ M81 data. Due to the large temporal coverage of the 
X-ray data, starting just six days after the explosion and comprising approximately one 
year of PSPC and five years of HRI observations, SN~1993J is one of the best observed SNe 
in the X-ray regime (Zimmermann et al. 1994a). The complete \R\ HRI+PSPC lightcurve of 
SN~1993J is presented in Fig.~\ref{f3} (source H18/P36). For the observed maximum 
PSPC count rate of $7.89\times 10^{-2}$ cts s$^{-1}$, an unabsorbed (0.5--2~keV band) 
flux of $f_{\rm x}=1.3\times 10^{-12}$ ergs cm$^{-2}$ s$^{-1}$ is obtained for an 
assumed thermal plasma with a temperature of $T=10^{6.5}$~K typical for young SNe. This 
converts to a peak X-ray luminosity of $L_{\rm x}=2.0\times 10^{39}$ ergs s$^{-1}$.
The PSPC data show $S_{\rm x} \propto t^{-\alpha}$ with $\alpha = (0.65 \pm 0.13)$,
consistent with the observed decline of radio emission
($S_{\lambda} \propto t^{-0.64}$ at $\lambda = 1.3, 2, 6$ and $20~{\rm cm}$;
van Dyk et al. 1994). Since both the radio and X-ray emissions are considered to be 
linked to the interaction of the SN blast wave with the ambient circumstellar medium,
similar results regarding the rate of decline are expected. 
An in-depth analysis of the X-ray characteristics and implications 
for the SN progenitor evolution will be presented in a separate publication.

\noindent $\bullet$ {\sl The dwarf companion Holmberg IX:} \\
H42/P63 coincides with the position of the Magellanic irregular dwarf galaxy Holmberg IX 
(DDO 66, UGC 05336;  R.A., Dec. (J2000) $= 09^{\rm h} 57^{\rm m} 30\fs1, 
+69^\circ 02^\prime 52^{\prime\prime}$; offset $3\farcs1$).
A revised distance estimation of Ho IX by Georgiev et al. \cite{Georgiev91}
shows that this dwarf galaxy is a close neighbor of M81 not only in projection
but also in space (maximum offset 500~kpc). Assuming a distance of 3.6~Mpc,
the inferred X-ray luminosity is $2.5 \times 10^{37}$ ergs s$^{-1}$.
A SNR was also detected in Ho IX (Hopp et al. 1996). The observed X-ray flux may be due to 
the presence of several unresolved young massive stars and/or relatively young SNRs.

\section{Diffuse X-ray Emission}
\label{results_diffuse}

The X-ray emission around the nucleus of M81 is clearly extended (Fig.~\ref{f4}), 
indicative of the presence of diffuse X-ray emission in the bulge region and/or
a combination of faint point-like sources. Excluding the two detected sources
H23 and H25 and subtracting the LMC X-1 profile, the total residual luminosity is 
$(3.8\pm0.2) \times 10^{39}$ ergs s$^{-1}$ (17\%) in the bulge region and 
$(7.0\pm0.3) \times 10^{39}$ ergs s$^{-1}$ (25\%) within the $D_{25}$ ellipse of M81
(Table~\ref{emission_components}). 

Our spectral analysis provides further evidence for multiple X-ray components 
in the central region of M81. A simple power-law fit to the PSPC spectrum,
extracted from a circle of $1'$ radius around the nucleus, is far from
being satisfactory (the reduced $\chi^2 \sim 10$). To constrain the potential presence
of other spectral components, we fix the power-law photon index at 1.85, 
as has been well determined by both \A\ and \B\ observations, which are sensitive to
photons in higher energy bands ($\gs2$~keV). We find that at least two more components are required
to provide a satisfactory fit to the PSPC spectrum. We choose 
optically thin plasma models to characterize these two components. Table~\ref{spectrum}
lists the spectral parameters obtained from the spectral fit (reduced $\chi^2 \ls 1$). 
The higher temperature component is typical for a combination of X-ray binaries and SNRs, 
as observed in other nearby galaxies (e.g., for the disk of NGC~253; Pietsch et al 2000). 
The lower temperature component indicates the presence of diffuse hot gas. The two thermal
components together account for $\sim13\%$ of the total M81 emission from the central region, 
in reasonably good agreement with the amount of the residual emission estimated
from the radial surface brightness profile after the nuclear source subtraction.
Our results are consistent with a recent examination of a \C\ M81 observation
(Tennant et al. 2000) which also indicates the presence of diffuse hot gas within 
the bulge, unaccounted for by weak unresolved sources.

\section{Comparison with M101}
\label{m101}

It is interesting to compare the X-ray properties of M81 and M101 (Fig.~\ref{f5}), 
for which we have carried out a similar analysis based on deep \R\ observations 
(Wang et al. 1999). While the former is an early-type spiral, the latter is a late-type 
(Sc) spiral with active star formation.
The X-ray emission from M101 is well correlated with spiral arms, especially
the active southeast arm, and is detected and resolved in the most luminous giant \hii\
regions (NGC 5447, 5455, 5461, and 5462). Although a fraction of the emission may be due
to X-ray binaries and to very young SNRs, hot gas is most likely an
important contributor. In general, there is a good correlation of diffuse X-ray emission
with tracers of recent massive star forming activities in M101 (e.g., H$\alpha$ intensity). 
In contrast, the M81 near-UV emission can be clearly decomposed
into an inner $<2'$ bulge and prominent spiral arm emission. The X-ray emission 
from M81 is not significantly enhanced along the spiral arms but more confined to the 
inner region of the galaxy.

Fig.~\ref{f6} further compares the mean radial intensity distributions of the diffuse 
(i.e. individual detected sources excised) X-ray emission with optical and UV intensity 
profiles of the galaxies. The optical profiles are significantly different from both the 
UV and the X-ray profiles for both galaxies.
The X-ray distribution of M101 is nearly identical to that of
the UV radiation and is substantially flatter than that of the optical light, suggesting 
that massive stars are responsible for much of the diffuse X-ray emission from the galaxy.
The X-ray distribution of M81, on the other hand, is much steeper at radii $\le2'$ compared 
to M101. The UV profile of M81 in general resembles the X-ray profile, apart from a clear 
excess at an off-center distance of $\sim2'$--$7'$. This excess is caused by an enhanced UV 
surface brightness in the spiral arms, especially along the northern spiral 
arm. But no enhanced diffuse X-ray emission is observed in these regions.
We find that the lack of enhanced X-ray emission associated with spiral arms in M81 can
be explained by the strong X-ray absorption by cool gas, as traced by \hi\ (Fig. 7).
The typical column density in the arms is several times $10^{21} {\rm~cm^{-2}}$, which
is sufficient to absorb the bulk of the X-ray emission from hot gas with a characteristic
temperature of $\sim0.2$~keV. Unlike M101, M81 contains no giant \hii\ regions that may 
produce hot superbubbles energetic enough to blow out from spiral arms. Furthermore,
M101 is a nearly face-on galaxy. So the line-of-sight absorption is significantly
smaller than in M81 and a better correlation between soft X-ray emission and spiral arms 
in M101 is expected. We thus conclude that the soft diffuse X-ray emission in M81 is 
also likely associated with recent massive star forming activities.

\section {Summary}
\label{summary}

We have systematically analyzed \R\ PSPC and HRI observations of the early-type 
spiral M81 to disentangle different emission components and to derive spatial, 
spectral and timing characteristics of X-ray sources in the field.
The main results and conclusions are as follows:

\noindent $\bullet$
Within a region of $17'$ radius around the M81 nucleus, 69 PSPC and 47 HRI sources 
are detected. Ten of them are found to be variable. Eleven HRI and nine additional 
PSPC sources are likely foreground or background objects (i.e. AGN, quasars, Galactic 
stars, etc.). Excluding these interlopers and the nucleus, 26 X-ray 
sources are within the $D_{25}$ ellipse of the galaxy and have luminosities 
in the range of $8\times10^{36}$  -- $1.2\times10^{39}$ ergs s$^{-1}$ in 
the 0.5--2~keV band. These sources account for $\sim10\%$ of the total luminosity 
($\sim3\times10^{40}$ ergs s$^{-1}$) of the galaxy. 

\noindent $\bullet$
We find an apparent association of luminous and variable X-ray sources with shock-heated
optical nebulae. These sources are in the luminosity range of 
$10^{37}$--$10^{40}~{\rm ergs~s}^{-1}$. Three position coincidences are with nebulae
that are previously classified as SNR candidates, and one with a nebula as a
superbubble. These associations suggest that such ISM structures may be powered by 
energetic outflows from X-ray binaries. 

\noindent $\bullet$
X-ray emission is also detected from Holmberg IX, the dwarf companion of M81.
Assuming a distance (3.6~Mpc) similar to that of M81, the inferred X-ray 
luminosity of Ho IX is $2.5 \times 10^{37}$ ergs s$^{-1}$, which can be easily 
account for by a number of relatively young SNRs and X-ray binaries.

\noindent $\bullet$
We present strong spatial and spectral evidence for the presence of an apparently 
diffuse X-ray component in M81. The diffuse X-ray emission is most prominent in the bulge 
region of the galaxy, although a fraction must arise in faint discrete stellar sources.
Diffuse X-ray emission fills much of the $D_{25}$ ellipse of M81, accounting for $\sim25\%$ 
($7\times10^{39}$ ergs s$^{-1}$) of the total X-ray luminosity of the galaxy.
A significant amount of X-ray emission may be absorbed by cool gas in the spiral arms.  
Nevertheless, the similarity between the radial X-ray and UV intensity profiles and 
a comparison with M101 suggest that the diffuse X-ray emission is 
associated with recent massive star activities in M81. 

\acknowledgments

We wish to thank M. Yun for providing the \hi\ map used in Fig. 7.
This research made use of various online services and databases,
e.g. ADS, HEASARC, NED, and NCSA ADIL as well as the ESO science archive facility 
and the \R\ data archive at MPE. The project is supported by NASA under the grants
 NAG 5-8999 and NAG5-9429.


\clearpage

\begin{deluxetable}{lrrrrrrrrrr}
\tabletypesize{\footnotesize}
\tablecaption{General parameters of M81 (NGC~3031) \label{parameters}}
\tablewidth{0pt}
\tablehead{
\colhead{Parameter} &
\colhead{Value} & 
\colhead{Ref.}}
\startdata
Type     \dotfill & SA(s)ab & 1 \\
         \dotfill & Sy1.8/LINER & 2 \\
Position of center (J2000) \dotfill & 
	R.A.~~~$09^{\rm h} 55^{\rm m} 33\fs2$ & 2 \\
\dotfill & Dec.~$+69\degr 03' 55\fs06$ & 2 \\
Distance \dotfill & 3.6~Mpc & 3 \\
\dotfill & ($1'~\cor~1$~kpc) & \\
Galactic foreground $N_{\rm H}$ \dotfill & 
$4.3 \times 10^{20}$~cm$^{-2}$ & 4 \\
Inclination \dotfill & $32\degr$ & 5 \\
Position angle of major axis \dotfill & $150\degr$ & 5 \\
Diameter \dotfill & $27' \times 14'$ & 2 \\
Blue Magnitude  \dotfill & 8 mag & 1
\enddata
\tablerefs{
(1) de Vaucouleurs et al. 1991;
(2) NED;
(3) Freedman et al. 1994;
(4) Dickey \& Lockman 1990;
(5) Garcia-Gomez \& Athanassoula 1991.
}
\end{deluxetable}

\begin{deluxetable}{lllc}
\tabletypesize{\footnotesize}
\tablecaption{\R\ Observations of M81 \label{obs_tab}}
\tablewidth{0pt}
\tablehead{
\colhead{Instr.} & 
\colhead{Sequence} & 
\colhead{Date} & 
\colhead{Obs.} \\
& \colhead{No.} & & \colhead{(ks)}}
\startdata
HRI & 600247h   &  1992 Oct 23 -- 27     & 26.6 \\
    & 600247h-1 &  1993 Apr 17 -- May 14 & 21.3 \\
    & 600739h   &  1994 Oct 19 -- 21     & 20.0 \\
    & 600740h   &  1995 Apr 13 -- May 4  & 19.2 \\
    & 600881h   &  1995 Oct 12 -- 25     & 14.9 \\
    & 600882h   &  1996 Apr 15 -- May 7  & 18.5 \\
    & 600882h-1 &  1996 Oct 27 -- Nov 10 & \phantom{0}5.1 \\
    & 601001h   &  1997 Mar 29 -- Apr 1  & 19.4 \\
    & 601002h   &  1997 Sep 30 -- Oct 16 & 19.8 \\
    & 601095h   &  1998 Mar 25 -- 26     & 12.6 \\
\noalign{\smallskip}
\tableline
\noalign{\smallskip}
PSPC & 600101p   &  1991 Mar 25 -- 27    & \phantom{0}9.6 \\
     & 600101p-1 &  1991 Oct 16 -- 17    & 11.5 \\
     & 600382p   &  1992 Sep 29 -- Oct 3 & 28.1 \\
     & 180015p   &  1993 Apr 3  -- 24    & 18.6 \\
     & 180015p-1 &  1993 May 4  -- 6     & \phantom{0}9.0 \\
     & 180035p   &  1993 Nov 1  -- 2     & 18.4 \\
     & 180035p-1 &  1993 Nov 7  -- 8     & \phantom{0}4.4 \\
     & 180050p   &  1994 Mar 31 -- Apr 2 & \phantom{0}1.9 
\enddata
\end{deluxetable}

\begin{deluxetable}{lrrrrrrrr}
\tabletypesize{\scriptsize}
\tablecaption{\R\ HRI M81 Source List \label{hri_sourcelist}}
\tablewidth{0pt}
\tablehead{
\colhead{Source} &
\colhead{R.A.$^{\rm 2000}$} &
\colhead{Dec.$^{\rm 2000}$} &
\colhead{R$_{\rm err}$} &
\colhead{Rate} &
\colhead{Comment} \\
&
\colhead{(h~~~m~~~s)} &
\colhead{($^{\circ}~~~'~~~''$)} &
\colhead{($''$)} &
$(10^{-4}~{\rm cts~s}^{-1})$ \\
\noalign{\medskip}
\colhead{(1)} &
\colhead{(2)} &
\colhead{(3)} &
\colhead{(4)} &
\colhead{(5)} &
\colhead{(6)}}
\startdata
%
%
H1 &
09~52~41.15 &
+69~02~44.0 &
$16.8$ &
$6.2$ $\pm$$
\phantom{0}1.5$  \\
%
%
H2 &
09~53~10.04 &
+69~00~07.6 &
$9.6$ &
$5.6$ $\pm$$
\phantom{0}1.2$  \\
%
%
H3 &
09~53~17.01 &
+69~06~46.0 &
$9.3$ &
$3.5$ $\pm$$
\phantom{0}0.9$ \\
%
%
H4 &
09~53~49.15 &
+68~52~42.2 &
$8.6$ &
$7.9$ $\pm$
$\phantom{0}1.4$ \\
%
%
H5 &
09~53~51.32 &
+69~02~47.6 &
$6.3$ &
$2.7$ $\pm$
$\phantom{0}0.7$ \\
%
%
H6 &
09~53~57.04 &
+69~03~57.0 &
$4.8$ &
$3.9$ $\pm$
$\phantom{0}0.7$ \\
%
%
H7 &
09~54~16.49 &
+69~16~27.8 &
$10.7$ &
$4.9$ $\pm$
$\phantom{0}1.2$ \\
%
%
H8 &
09~54~45.27 &
+68~57~00.1 &
$3.9$ &
$10.7$ $\pm$
$\phantom{0}1.0$  \\
%
%
H9 &
09~54~50.96 &
+69~02~53.5 &
$4.2$ &
$2.4$ $\pm$
$\phantom{0}0.5$ & var \\
%
%
H10 &
09~55~00.06 &
+69~07~48.4 &
$3.7$ &
$9.2$ $\pm$
$\phantom{0}0.8$ \\
%
%
H11 &
09~55~01.87 &
+68~56~22.1 &
$4.0$ &
$11.7$ $\pm$
$\phantom{0}1.0$ \\
%
%
H12 &
09~55~02.43 &
+68~50~50.9 &
$12.7$ &
$4.8$ $\pm$
$\phantom{0}1.2$ \\
%
%
H13 &
09~55~09.74 &
+69~04~10.1 &
$3.7$ &
$7.5$ $\pm$
$\phantom{0}0.7$ & var \\
%
%
H14 &
09~55~10.27 &
+69~05~04.7 &
$3.6$ &
$13.1$ $\pm$
$\phantom{0}0.9$ \\
%
%
H15 &
09~55~22.04 &
+69~05~14.0 &
$3.8$ &
$8.9$ $\pm$
$\phantom{0}0.8$ \\
%
%
H16 &
09~55~22.03 &
+69~06~41.2 &
$4.2$ &
$1.7$ $\pm$
$\phantom{0}0.4$ \\
%
%
H17 &
09~55~24.37 &
+69~10~02.0 &
$3.5$ &
$34.9$ $\pm$
$\phantom{0}1.5$ & var \\
%
%
H18 &
09~55~24.72 &
+69~01~15.0 &
$3.5$ &
$79.4$ $\pm$
$\phantom{0}2.2$ & var \\
%
%
H19 &
09~55~25.78 &
+69~15~49.8 &
$10.1$ &
$3.1$ $\pm$
$\phantom{0}0.9$ \\
%
%
H20 &
09~55~28.86 &
+69~06~16.1 &
$4.9$ &
$1.2$ $\pm$
$\phantom{0}0.4$ \\
%
%
H21 &
09~55~33.05 &
+69~00~34.9 &
$3.5$ &
$200.3$ $\pm$
$\phantom{0}3.4$ & var \\
%
%
H22 &
09~55~33.32 &
+69~03~57.5 &
$3.5$ &
$2948.5$ $\pm$
$12.9$ & var \\
%
%
H23 &
09~55~35.06 &
+69~03~20.6 &
$3.8$ &
$18.5$ $\pm$
$\phantom{0}1.2$ & \\
%
%
H24 &
09~55~35.08 &
+68~55~15.7 &
$6.3$ &
$2.5$ $\pm$
$\phantom{0}0.6$ \\
%
%
H25 &
09~55~42.13 &
+69~03~39.4 &
$3.7$ &
$18.4$ $\pm$
$\phantom{0}1.2$ & var \\
%
%
H26 &
09~55~43.60 &
+69~17~01.8 &
$9.7$ &
$5.9$ $\pm$
$\phantom{0}1.2$ \\
%
%
H27 &
09~55~43.95 &
+68~59~07.3 &
$4.0$ &
$2.4$ $\pm$
$\phantom{0}0.5$ \\
%
%
H28 &
09~55~47.44 &
+69~05~54.5 &
$4.1$ &
$3.7$ $\pm$
$\phantom{0}0.6$ \\
%
%
H29 &
09~55~49.42 &
+68~58~38.0 &
$3.9$ &
$5.6$ $\pm$
$\phantom{0}0.7$ \\
%
%
H30 &
09~55~49.54 &
+69~08~16.0 &
$4.0$ &
$3.0$ $\pm$
$\phantom{0}0.5$ \\
%
%
H31 &
09~55~50.10 &
+69~05~34.8 &
$3.6$ &
$19.6$ $\pm$
$\phantom{0}1.1$ \\
%
%
H32 &
09~55~58.79 &
+69~05~29.7 &
$4.1$ &
$2.6$ $\pm$
$\phantom{0}0.5$ \\
%
%
H33 &
09~56~02.23 &
+68~59~01.0 &
$4.3$ &
$3.1$ $\pm$
$\phantom{0}0.6$ \\
%
%
H34 &
09~56~02.90 &
+68~59~36.7 &
$4.5$ &
$2.2$ $\pm$
$\phantom{0}0.5$ \\
%
%
H35 &
09~56~09.59 &
+69~12~53.4 &
$5.5$ &
$5.0$ $\pm$
$\phantom{0}0.8$ \\
%
%
H36 &
09~56~09.22 &
+69~01~09.6 &
$3.7$ &
$8.9$ $\pm$
$\phantom{0}0.8$ \\
%
%
H37 &
09~56~13.85 &
+69~06~34.0 &
$4.4$ &
$2.3$ $\pm$
$\phantom{0}0.5$ \\
%
%
H38 &
09~56~14.19 &
+68~57~24.8 &
$3.7$ &
$10.5$ $\pm$
$\phantom{0}0.9$ & var  \\
%
%
H39 &
09~56~36.87 &
+69~00~30.1 &
$3.8$ &
$7.3$ $\pm$
$\phantom{0}0.8$ \\
%
%
H40 &
09~57~01.58 &
+68~55~01.3 &
$4.3$ &
$16.3$ $\pm$
$\phantom{0}1.3$ \\
%
%
H41 &
09~57~11.37 &
+69~05~04.1 &
$5.3$ &
$2.7$ $\pm$
$\phantom{0}0.6$ \\
%
%
H42 &
09~57~31.26 &
+69~02~31.9 &
$12.3$ &
$4.1$ $\pm$
$\phantom{0}1.0$ \\
%
%
H43 &
09~57~35.80 &
+69~00~09.1 &
$7.0$ &
$7.5$ $\pm$
$\phantom{0}1.1$ \\
%
%
H44 &
09~57~53.76 &
+69~03~50.3 &
$3.5$ &
$552.9$ $\pm$
$\phantom{0}5.9$ & var \\
%
%
H45 &
09~57~56.44 &
+69~11~39.2 &
$12.6$ &
$6.9$ $\pm$
$\phantom{0}1.4$ \\
%
%
H46 &
09~58~02.94 &
+68~57~10.1 &
$5.4$ &
$22.7$ $\pm$
$\phantom{0}1.8$ \\
\enddata
\end{deluxetable}

\begin{deluxetable}{lrrrrrrrrr}
\tabletypesize{\scriptsize}
\tablecaption{\R\ PSPC M81 Source List \label{pspc_sourcelist}}
\tablewidth{0pt}
\tablehead{
\colhead{Source} &
\colhead{R.A.$^{\rm 2000}$} &
\colhead{Dec.$^{\rm 2000}$} &
\colhead{Rate} &
\multicolumn{2}{c}{Hardness Ratio} &
\colhead{Comment} \\
& \colhead{(h~~~m~~~s)} &
\colhead{($^{\circ}~~~'~~~''$)} &
\colhead{(cts~ks$^{-1}$)} &
\colhead{HR1} &
\colhead{HR2} & \\
\noalign{\medskip}
\colhead{(1)} &
\colhead{(2)} &
\colhead{(3)} &
\colhead{(4)} &
\colhead{(5)} &
\colhead{(6)} &
\colhead{(7)}}
\startdata
P1 &  09 52 40.9 & +69 03 50 &      0.8 $\pm$ 0.1 &   0.89 $\pm$ 0.28 &   0.73 $\pm$ 0.19 & \\ 
P2 &  09 52 44.0 & +69 02 52 &      0.7 $\pm$ 0.1 &   1.00 $\pm$ 0.36 &   0.62 $\pm$ 0.21 & H1 \\ 
P3 &  09 52 47.4 & +69 09 01 &      0.7 $\pm$ 0.1 &   1.00 $\pm$ 0.36 &   0.39 $\pm$ 0.22 & \\ 
P4 &  09 52 50.7 & +68 59 21 &      0.5 $\pm$ 0.1 &   0.79 $\pm$ 0.41 &   0.06 $\pm$ 0.28 & \\ 
P5 &  09 53 00.1 & +69 07 27 &      0.5 $\pm$ 0.1 &   0.80 $\pm$ 0.36 &   0.56 $\pm$ 0.24 & \\ 
P6 &  09 53 11.7 & +68 59 59 &      1.0 $\pm$ 0.1 &   1.00 $\pm$ 0.21 &   0.18 $\pm$ 0.14 & H2 \\ 
P7 &  09 53 18.1 & +69 06 37 &      0.9 $\pm$ 0.1 &   0.56 $\pm$ 0.14 &   0.42 $\pm$ 0.12 & H3 \\ 
P8 &  09 53 27.4 & +69 04 13 &      0.9 $\pm$ 0.1 &   1.00 $\pm$ 0.18 &   0.55 $\pm$ 0.12 & \\ 
P9 &  09 53 29.8 & +68 58 35 &      1.8 $\pm$ 0.2 &   0.33 $\pm$ 0.08 &   0.23 $\pm$ 0.10 & \\ 
P10 &  09 53 37.0 & +69 05 36 &     0.5 $\pm$ 0.1 &   0.96 $\pm$ 0.35 &   0.27 $\pm$ 0.22 & \\ 
P11 &  09 53 40.8 & +68 59 16 &     1.4 $\pm$ 0.2 &   0.87 $\pm$ 0.12 &   0.59 $\pm$ 0.10 & \\ 
P12 &  09 53 43.6 & +69 16 00 &     0.5 $\pm$ 0.1 &   1.00 $\pm$ 0.53 &   0.51 $\pm$ 0.32 & \\ 
P13 &  09 53 50.2 & +68 52 39 &     3.3 $\pm$ 0.2 &   0.84 $\pm$ 0.06 &   0.42 $\pm$ 0.07 & H4 \\ 
P14 &  09 53 52.2 & +69 02 51 &     1.1 $\pm$ 0.1 &   1.00 $\pm$ 0.15 &   0.35 $\pm$ 0.12 & H5 \\ 
P15 &  09 53 58.0 & +69 03 56 &     1.3 $\pm$ 0.1 &   0.62 $\pm$ 0.12 &   0.28 $\pm$ 0.12 & H6 \\ 
P16 &  09 54 21.0 & +69 00 03 &     1.0 $\pm$ 0.1 &   0.90 $\pm$ 0.14 &   0.47 $\pm$ 0.12 & \\ 
P17 &  09 54 21.7 & +68 54 36 &     1.7 $\pm$ 0.2 &   0.90 $\pm$ 0.13 &   0.23 $\pm$ 0.11 & \\ 
P18 &  09 54 31.8 & +68 52 36 &     0.4 $\pm$ 0.1 &   0.21 $\pm$ 0.24 &   0.45 $\pm$ 0.27 & \\ 
P19 &  09 54 39.3 & +69 19 13 &     0.8 $\pm$ 0.2 &   1.00 $\pm$ 0.29 &   0.09 $\pm$ 0.19 & \\ 
P20 &  09 54 41.3 & +69 04 51 &     0.7 $\pm$ 0.1 &   0.97 $\pm$ 0.25 &   0.26 $\pm$ 0.18 & \\ 
P21 &  09 54 42.6 & +69 02 38 &     0.4 $\pm$ 0.1 &   0.39 $\pm$ 0.24 &   0.46 $\pm$ 0.24 & H9; var \\ 
P22 &  09 54 45.3 & +68 56 58 &     3.3 $\pm$ 0.2 &   0.90 $\pm$ 0.05 &   0.10 $\pm$ 0.06 & H8 \\ 
P23 &  09 54 47.6 & +69 11 22 &     0.4 $\pm$ 0.1 &   1.00 $\pm$ 0.43 &   0.19 $\pm$ 0.25 & \\ 
P24 &  09 55 00.2 & +69 19 14 &     0.7 $\pm$ 0.1 &   0.78 $\pm$ 0.30 &   0.11 $\pm$ 0.22 & \\ 
P25 &  09 55 00.8 & +69 07 40 &     4.4 $\pm$ 0.2 &   0.95 $\pm$ 0.04 &   0.51 $\pm$ 0.05 & H10 \\ 
P26 &  09 55 02.4 & +69 10 29 &     0.7 $\pm$ 0.1 &   1.00 $\pm$ 0.26 &   0.37 $\pm$ 0.16 & \\ 
P27 &  09 55 02.4 & +68 56 21 &     1.7 $\pm$ 0.2 & $-0.09$ $\pm$ 0.06 &  $-0.06$ $\pm$ 0.10 & H11 \\ 
P28 &  09 55 05.5 & +68 58 53 &     0.3 $\pm$ 0.1 &   1.00 $\pm$ 0.44 &   0.06 $\pm$ 0.28 & \\ 
P29 &  09 55 10.5 & +69 04 04 &     5.1 $\pm$ 0.3 &   0.92 $\pm$ 0.03 &   0.35 $\pm$ 0.05 & H13; var \\ 
P30 &  09 55 11.0 & +69 08 29 &     0.6 $\pm$ 0.1 &   1.00 $\pm$ 0.27 &   0.25 $\pm$ 0.19 & \\ 
P31 &  09 55 11.0 & +69 04 60 &     5.6 $\pm$ 0.3 &   0.99 $\pm$ 0.03 &   0.48 $\pm$ 0.04 & H14 \\ 
P32 &  09 55 13.6 & +69 12 32 &     0.4 $\pm$ 0.1 &   0.74 $\pm$ 0.38 &   0.85 $\pm$ 0.27 & \\ 
P33 &  09 55 22.7 & +69 06 31 &     7.3 $\pm$ 0.3 &   0.89 $\pm$ 0.03 &   0.43 $\pm$ 0.04 & \\ 
P34 &  09 55 23.7 & +69 05 08 &     6.1 $\pm$ 0.3 &   0.91 $\pm$ 0.03 &   0.44 $\pm$ 0.04 & H15 \\ 
P35 &  09 55 24.6 & +69 09 54 &    12.3 $\pm$ 0.4 &   0.90 $\pm$ 0.02 &   0.28 $\pm$ 0.03 & H17; var \\ 
P36 &  09 55 24.9 & +69 01 13 &    27.0 $\pm$ 0.6 &   0.96 $\pm$ 0.01 &   0.43 $\pm$ 0.02 & H18; var\\ 
P37 &  09 55 33.1 & +69 00 34 &    70.1 $\pm$ 0.9 &   0.98 $\pm$ 0.00 &   0.54 $\pm$ 0.01 & H21; var \\ 
P38 &  09 55 33.6 & +69 03 54 &   666.5 $\pm$ 2.8 &   0.86 $\pm$ 0.00 &   0.26 $\pm$ 0.00 & H22; var\\ 
P39 &  09 55 41.0 & +69 17 21 &    1.1 $\pm$ 0.2 &   0.60 $\pm$ 0.15 &   0.52 $\pm$ 0.13 & H26 \\ 
P40 &  09 55 49.2 & +68 58 41 &    2.2 $\pm$ 0.2 &   0.90 $\pm$ 0.07 &   0.23 $\pm$ 0.08 & \\ 
P41 &  09 55 49.4 & +69 05 32 &    8.1 $\pm$ 0.3 &   0.86 $\pm$ 0.02 &   0.43 $\pm$ 0.04 & H28+H31 \\ 
P42 &  09 55 50.0 & +69 08 09 &    1.5 $\pm$ 0.2 &   0.91 $\pm$ 0.09 &   0.20 $\pm$ 0.09 & H30 \\ 
P43 &  09 56 01.5 & +68 59 09 &    1.8 $\pm$ 0.2 &   0.86 $\pm$ 0.10 &   0.20 $\pm$ 0.10 & H33+H34 \\ 
P44 &  09 56 08.9 & +69 01 07 &    1.5 $\pm$ 0.1 &   0.81 $\pm$ 0.10 & $-0.60$ $\pm$ 0.10 & H36 \\ 
P45 &  09 56 09.2 & +69 12 45 &    1.0 $\pm$ 0.1 &   1.00 $\pm$ 0.17 &   0.18 $\pm$ 0.14 & H35 \\ 
P46 &  09 56 12.7 & +69 06 29 &    0.4 $\pm$ 0.1 &   1.00 $\pm$ 0.37 &   0.35 $\pm$ 0.22 & H37 \\ 
P47 &  09 56 14.2 & +68 57 27 &   2.2 $\pm$ 0.2 &   0.39 $\pm$ 0.07 &   0.07 $\pm$ 0.08 & H38; var \\ 
P48 &  09 56 36.3 & +69 00 28 &   2.8 $\pm$ 0.2 &   0.86 $\pm$ 0.06 &   0.31 $\pm$ 0.07 & H39 \\ 
P49 &  09 56 43.7 & +68 53 51 &   1.8 $\pm$ 0.1 &   0.87 $\pm$ 0.12 &   0.54 $\pm$ 0.10 & \\ 
P50 &  09 56 46.0 & +68 54 40 &   1.9 $\pm$ 0.2 &   1.00 $\pm$ 0.10 &   0.58 $\pm$ 0.09 & \\ 
P51 &  09 56 52.0 & +69 07 41 &   0.5 $\pm$ 0.1 &   0.78 $\pm$ 0.26 &   0.36 $\pm$ 0.20 & \\ 
P52 &  09 56 52.6 & +69 10 42 &   0.9 $\pm$ 0.1 &   0.76 $\pm$ 0.16 &   0.09 $\pm$ 0.15 & \\ 
P53 &  09 56 52.9 & +69 11 59 &   0.8 $\pm$ 0.1 &   1.00 $\pm$ 0.20 &   0.49 $\pm$ 0.15 & \\ 
P54 &  09 56 55.8 & +69 08 60 &   0.4 $\pm$ 0.1 &   0.78 $\pm$ 0.33 &   0.47 $\pm$ 0.23 & \\ 
P55 &  09 57 01.4 & +68 54 60 &   4.4 $\pm$ 0.3 &   0.88 $\pm$ 0.04 &   0.12 $\pm$ 0.06 & H40 \\ 
P56 &  09 57 01.6 & +68 56 49 &   0.6 $\pm$ 0.1 &   0.93 $\pm$ 0.32 &   0.14 $\pm$ 0.23 & \\
P57 &  09 57 10.9 & +69 05 01 &   0.6 $\pm$ 0.1 &   1.00 $\pm$ 0.27 &   0.13 $\pm$ 0.20 & H41 \\ 
P58 &  09 57 14.4 & +69 11 35 &   0.5 $\pm$ 0.1 &   0.94 $\pm$ 0.29 &   0.21 $\pm$ 0.21 & \\ 
P59 &  09 57 17.2 & +69 10 12 &   0.5 $\pm$ 0.1 &   0.94 $\pm$ 0.30 &   0.48 $\pm$ 0.21 & \\ 
P60 &  09 57 17.5 & +68 58 27 &   1.4 $\pm$ 0.2 &   0.95 $\pm$ 0.12 &   0.33 $\pm$ 0.11 & \\ 
P61 &  09 57 26.9 & +68 53 16 &   0.6 $\pm$ 0.1 &   1.00 $\pm$ 0.41 &   0.08 $\pm$ 0.25 & \\ 
P62 &  09 57 28.4 & +69 13 30 &   1.8 $\pm$ 0.2 &   0.99 $\pm$ 0.11 &   0.25 $\pm$ 0.10 & \\ 
P63 &  09 57 29.8 & +69 02 32 &   0.8 $\pm$ 0.1 &   0.80 $\pm$ 0.18 &   0.02 $\pm$ 0.16 & H42 \\ 
P64 &  09 57 35.6 & +69 00 09 &   1.6 $\pm$ 0.2 &   1.00 $\pm$ 0.11 &   0.11 $\pm$ 0.10 & H43 \\
P65 &  09 57 35.7 & +69 16 07 &   0.7 $\pm$ 0.2 &   0.98 $\pm$ 0.27 &   0.37 $\pm$ 0.18 & \\
P66 &  09 57 53.3 & +69 03 48 & 198.0 $\pm$ 1.6 &   0.97 $\pm$ 0.00 &   0.39 $\pm$ 0.01 & H44; var \\ 
P67 &  09 57 55.7 & +69 11 33 &   3.1 $\pm$ 0.2 &   0.84 $\pm$ 0.07 &   0.22 $\pm$ 0.07 & H45 \\ 
P68 &  09 57 57.3 & +69 06 11 &    0.9 $\pm$ 0.2 &   0.70 $\pm$ 0.21 &   0.24 $\pm$ 0.20 & \\ 
P69 &  09 58 02.2 & +68 57 10 &   3.7 $\pm$ 0.3 &   0.16 $\pm$ 0.05 &   0.03 $\pm$ 0.07 & H46 \\ 
\enddata
\end{deluxetable}

\begin{deluxetable}{lrrrrrrr}
\tabletypesize{\scriptsize}
\tablecaption{\R\ M81 Source Identifications \label{ident}}
\tablewidth{0pt}
\tablehead{
\colhead{Source} &
\colhead{Offset} &
\colhead{$R$} &
\colhead{$B$} &
\colhead{$B-R$} &
\colhead{Identification\tablenotemark{a}} \\
& \colhead{($''$)} &
\colhead{(mag)} &
\colhead{(mag)} &
\colhead{(mag)} \\
\noalign{\medskip}
\colhead{(1)} &
\colhead{(2)} &
\colhead{(3)} &
\colhead{(4)} &
\colhead{(5)} &
\colhead{(6)}}
\startdata
H2 = P6 & 6.0 & 11.82 & 14.13 & \phantom{0}\phantom{0}\phantom{0}2.31   
	 & GS 0438301079; U1575 03021793 \\
H3 = P7  & 5.4 & 18.0 & 19.40 & \phantom{0}\phantom{0}\phantom{0}$1.40$
	 & U1575 03022008 \\
P9 & 12.0 & 11.02 & 12.47 & \phantom{0}\phantom{0}\phantom{0}1.72
	& GS 0438300727; U1575 03022347 \\
P11 & 6.9  & 19.27 & -- &  \phantom{0}$>1.92$   & non-stellar \\
H4 = P13 & 3.6 & 19.06 & 20.69 &  \phantom{0}\phantom{0}\phantom{0}1.63  
	 & U1575 03022935 \\
P14  & 12.6 & 18.17 & 21.18 &  \phantom{0}\phantom{0}\phantom{0}3.01  & non-stellar \\
H6 = P15 & 4.2 & 19.77 & -- &  \phantom{0}$>1.42$  
	 & stellar \\
P19  & 3.1 & 19.81 & -- &  \phantom{0}$>1.38$   & U1575 03024392 \\
H8 = P22 & 1.8 & 12.45 & 13.55 & \phantom{0}\phantom{0}\phantom{0}1.10  
	 & GS 0438300613; U1575 03024587 \\
H9  & 5.0 & -- & -- & --  & SNR MF4 \\
P23  & 7.4 & 18.90 & 19.50 &  \phantom{0}\phantom{0}\phantom{0}0.60  & U1575 03024612 \\
H11 = P27 & 5.9 & 10.31 & 11.62 &  \phantom{0}\phantom{0}\phantom{0}1.31
 	 & GS 0438301127; U1575 03025086 \\
H13 = P29 & 4.5 & -- & -- & -- & X2; SNR MF11 \\
H14 = P31 & 2.9 & -- & -- & -- & X3 \\
H17 = P35 & 6.3 & -- & -- & -- & X4 \\
H18 = P36 & 1.3 & -- & -- & -- & SN~1993J \\
H21 = P37 & 2.7 & -- & -- & -- & X6; SNR MF22 \\
H22 = P38 & 3.3 & -- & -- & -- & X5; M81 nucleus \\
H31       & 5.2 & -- & -- & -- & X7 \\
H38 = P47 & 0.5 & 16.85 & 19.37 &  \phantom{0}\phantom{0}\phantom{0}2.52 
	& X8; U1575 03027313 \\
H39 = P48 & 2.3 & 17.07 & 17.24 &  \phantom{0}\phantom{0}\phantom{0}0.17  
	& U1575 03027973 \\
P49 & 15.3 & 14.72 & 15.93 & \phantom{0}\phantom{0}\phantom{0}1.21  & GS 0438300776; U1575 03028257 \\
H40 = P55 & 0.7 & 16.91 & 17.37 &  \phantom{0}\phantom{0}\phantom{0}0.46 
	& U1575 03028660 \\
P56 & 5.0 & 18.71 & 20.48 &  \phantom{0}\phantom{0}\phantom{0}1.87 & U1575 03028666 \\
P58 & 8.0 & 14.84 & 16.09 &  \phantom{0}\phantom{0}\phantom{0}1.25 & GS 0438300612 \\
P62 & 10.6 & 18.51 & -- & \phantom{0}$>2.68$   & non-stellar \\
H42 = P63 & 3.1 & -- & 19.92 & $<-0.08$ & Ho IX \\
H44 = P66 & 4.3 & 17.80 & --  &  \phantom{0}$>3.39$ & X9; U1575 03030069; nebula \\
H45 = P67 & 8.5 & 19.44 & 21.17 &  \phantom{0}\phantom{0}\phantom{0}1.73 & non-stellar \\
H46 = P69 & 2.4 & \phantom{0}9.95 & 11.51 &  \phantom{0}\phantom{0}\phantom{0}1.56 & GS 0438301132 \\
\noalign{\smallskip}
\enddata
\tablenotetext{(a)\phantom{0}}{see \S\ref{point_sources} for details. Identifications are denoted by: \\
GS -- HST Guide Star Catalog v1.1; \\
U -- USNO A-v2.0 Catalog of Astrometric Standards; \\ 
MF -- SNR candidate number from Matonick \& Fesen \cite{Matonick97}; \\
X -- \E\ X-ray source number from Fabbiano \cite{Fabbiano88}; \\
Ho IX -- Holmberg IX dwarf galaxy.}
\end{deluxetable}

\begin{deluxetable}{lrrcrccccr}
\tabletypesize{\footnotesize}
\tablecaption{Position Coincidence between X-ray Sources and SNR Candidates \label{snr}}
\tablewidth{0pt}
\tablehead{
\colhead{SNR} &
\colhead{R.A.} &
\colhead{Dec.} &
\colhead{$\Delta_{\rm x-o}$} &
\colhead{$L_{{\rm H}\alpha}$} &
\colhead{[S{\sc ii}]/H$\alpha$} &
\colhead{Size} &
\colhead{Optical} &
\colhead{\R} &
\colhead{$L_{\rm x}$} \\
\colhead{Name} &
\colhead{(h~~~m~~~s)} &
\colhead{($^{\circ}~~~'~~~''$)} &
\colhead{($''$)} & 
\colhead{(ergs s$^{-1}$)} & &
\colhead{(pc)} &
\colhead{Shape} &
\colhead{Source} &
\colhead{(ergs s$^{-1}$})}
\startdata
MF4  & $09~54~51.3$ & $+69~02~58.5$ & 6.2 & $4.5\times10^{36}$ & 0.58 & 40 & stellar 
	& H9  &  $1.5\times10^{37}$ \\
MF11 & $09~55~09.6$ & $+69~04~14.6$ & 4.5 & $2.2\times10^{36}$ & 1.58 & 40 & stellar 
	& H13/P29 & $4.7\times10^{37}$ \\
MF22 & $09~55~32.7$ & $+69~00~32.9$ & 2.7 & $1.7\times10^{37}$ & 0.85 & 90 & filled  
	& H21/P37 & $1.2\times10^{39}$ \\
\noalign{\smallskip}
\enddata
\end{deluxetable}

\begin{deluxetable}{lccc}
\tabletypesize{\footnotesize}
\tablecaption{X-Ray Emission Components in the Field of M81 \label{emission_components}}
\tablewidth{0pt}
\tablehead{
\colhead{Emission component} &
\colhead{HRI Rate} & 
\colhead{$L_{\rm x}$\tablenotemark{a}} & 
\colhead{Fraction} \\ 
\colhead{} &
\colhead{(${\rm cts~s}^{-1}$)} & 
\colhead{($10^{40}~{\rm ergs~s}^{-1}$)} &
\colhead{(\%)}}
\startdata
Total Galaxy \dotfill & 
		0.457 & 2.83 & 100 \\
Point-like nuclear source (H22) \dotfill &
                0.295 & 1.83 & \phantom{0}65 \\
Interlopers (H11, H39, H40, H41) \dotfill &
		0.005 & 0.03 & \phantom{0}\phantom{0}1 \\
Point sources\tablenotemark{b} \dotfill &
		0.044 & 0.27 & \phantom{0}10 \\
Diffuse emission within the $D_{25}$ ellipse \dotfill & 
                0.113 & 0.70 & \phantom{0}24 \\
                \noalign{\smallskip}
                \hline
                \noalign{\smallskip}
Total bulge emission ($<2'$ radius) \dotfill & 
                 0.354 & 2.19 & 100 \\
Point-like nuclear source ($<2'$ radius) \dotfill &
                 0.292 & 1.81 & \phantom{0}83 \\
Extended bulge emission ($<2'$ radius)\tablenotemark{c} \dotfill & 
                 0.062 & 0.38 & \phantom{0}17 \\
\enddata
\tablenotetext{(a)\phantom{0}}{0.5--2~keV band luminosities.}
\tablenotetext{(b)\phantom{0}}{Point sources within the $D_{25}$ ellipse of M81
	(Table~\ref{hri_sourcelist}), excluding identified interlopers and the nuclear X-ray source.}
\tablenotetext{(c)\phantom{0}}{Derived from the radial intensity profile 
	of the point-like X-ray source Her X-1, scaled to the peak of the X-ray emission
	from the bulge (Fig.~\ref{f4}; \S\ref{diffuse}).}
\end{deluxetable}

\begin{deluxetable}{lccc}
\tabletypesize{\footnotesize}
\tablecaption{X-ray Spectral Properties of the Bulge Region \label{spectrum}}
\tablewidth{0pt}
\tablehead{
\colhead{Model\tablenotemark{a}} & 
\colhead{$T,\Gamma$\tablenotemark{b}} &
\colhead{$N_{\rm H}$\tablenotemark{b}} &
\colhead{$L_{\rm x}$\tablenotemark{c}} \\
& \colhead{(keV)} & 
\colhead{($10^{20}$ cm$^{-2}$)} &
\colhead{(ergs s$^{-1}$)}
}
\startdata
\multicolumn{1}{l}{Three-component model:} \\
\noalign{\smallskip}
\phantom{0}$\bullet$~Thermal Plasma 1 & 0.15 (0.13--0.17) & 4.8 (4.7--4.9) & 
	$2.0\times10^{39}$ \\
\phantom{0}$\bullet$~Thermal Plasma 2 & 0.63 (0.52--0.74) & 6.0 (5.7--6.3) & 
	$0.5\times10^{39}$ \\
\phantom{0}$\bullet$~Power Law        & 1.85 (1.83--1.87) & 23 (22--25)    & 
	$1.7\times10^{40}$ \\
\enddata
\tablenotetext{(a)}{\phantom{0} Metal abundances for both X-ray-emitting
 and -absorbing materials are assumed as 100\% solar.}
\tablenotetext{(b)}{\phantom{0} 90\% confidence intervals.}
\tablenotetext{(c)}{\phantom{0} 0.5--2~keV band luminosities.}
\end{deluxetable}


\clearpage

\figcaption[f1.ps]
{\R\ PSPC (0.5--2~keV band) contour map of M81,
overlayed onto a digitized DSS2 plate. The PSPC intensity map was adaptively 
smoothed with a Gaussian with size adjusted to achieve a constant signal-to-noise 
ratio of 6. Contours are at 4.5, 6, 10, 20, 50, 100, 500, 2000 and 5000 
$\times 10^{-3}$ cts arcmin$^{-2}$ s$^{-1}$.
\label{f1}}

\figcaption[f2.ps]
{\R\ PSPC (0.5--2~keV band) intensity contour map of M81. The map was smoothed with a 
Gaussian filter of $35''$ (FWHM). Same X-ray contour levels as in Fig.~\ref{f1}.
HRI and PSPC source positions are marked with crosses and boxes, respectively, and are
enumerated according to the source lists (Tables~\ref{hri_sourcelist} and \ref{pspc_sourcelist})
in the right-hand panels. The $D_{25}$ ellipse of M81 is indicated by a dotted line.
The lower panel shows a high-resolution HRI  close-up of the inner
$11' \times 11'$ region. The map was smoothed with a Gaussian filter of $9''$ (FWHM). 
Contours are at 2, 4, 6, 10, 20, 50, 100, 500, 2000 and 5000 
$\times 10^{-3}$ cts arcmin$^{-2}$ s$^{-1}$.
\label{f2}}

\figcaption[f3.ps]
{Lightcurves of the five brightest X-ray sources in the
M81 field. PSPC data are marked as boxes, HRI data are indicated by a horizontal bar.
Dashed lines give the mean count rates over the \R\ PSPC period of observation,
the dotted lines indicate the mean count rates over the complete \R\ HRI observation.
Error bars are $1\sigma$ statistical errors, arrows indicate $3\sigma$ upper limits.
\label{f3}}

\figcaption[f4.ps]
{Radial X-ray surface brightness profile of the M81 bulge region (solid line),
compared to the profile of the point-like X-ray source LMC X-1 (dashed line), which is
scaled to the peak of the bulge emission.
\label{f4}}

\figcaption[f5a.ps,f5b.ps]
{\R\ HRI intensity contours of M81 (left-hand panel) and M101 (right-hand panel), 
overlaid on UIT UV images. The HRI images are adaptively smoothed with a S/N equal to 6. 
Contour levels are at 0.4, 0.6, 1.0, 1.5, 2.3, 3.3., 5.3, 11, 23, 47, 80, 160, 300, 600, 
1200, and 2400 $\times 10^{-3}$ cts arcmin$^{-2}$ s$^{-1}$.
\label{f5}}

\figcaption[f6a.ps,f6b.ps]
{Radial diffuse X-ray intensity distributions of M81 (left-hand panel)
and M101 (right-hand panel), compared with UV (solid line) and optical 
(dashed line) intensity profiles around the galaxies nuclei. 
The intensity units are for X-ray data only; the optical (POSS E survey) and 
UV (Astro-1 UIT) profiles are arbitrarily normalized.
\label{f6}}

\figcaption[f7.ps]
{\hi\ contours overlaid on the near-UV image of M81. The \hi\ data are the same as
in Yun et al. (1994) and the contours are at 5, 10, 15, 20, and 25 
$\times 10^{20} {\rm~cm^{-2}}$.
\label{f7}}

\end{document}